# The use of 5G Non-Public Networks to support Industry 4.0 scenarios

Jose Ordonez-Lucena, Jesús Folgueira Chavarria, Luis M. Contreras, Antonio Pastor
Telefonica I+D, Global CTIO, Madrid, Spain
Email: {joseantonio.ordonezlucena, jesus.folgueira, luismiguel.contrerasmurillo, antonio.pastorperales}@telefonica.com

*Abstract*—The on-going digital transformation is key to progress towards a new generation of more efficient, sustainable and connected industrial systems allowing the so-called factories of the future. This new generation, commonly referred to as industry 4.0, will be accompanied by a new wave of use cases that will allow companies from logistics and manufacturing sectors to increase flexibility, productivity and usability in the industrial processes executed within their factory premises. Unlike typical use cases from other vertical sectors (e.g. energy, media, smart cities), industry 4.0 use cases will bring very stringent requirements in terms of latency, reliability and high-accuracy positioning. The combination of 5G technology with enterprise network solutions becomes crucial to satisfy these requirements in indoor, private environments. In this context, the concept of 5G non-public networks has emerged. In this article we provide an overview of 5G non-public networks, studying their applicability to the industry 4.0 ecosystem. On the basis of the work (being) developed in 3GPP Rel-16 specifications, we identify a number of deployment options relevant for non-public networks, and discuss their integration with mobile network operators' public networks. Finally, we provide a comparative analysis of these options, assessing their feasibility according to different criteria, including technical, regulatory and business aspects. The outcome of this analysis will help industry players interested in using non-public networks to decide which is the most appropriate deployment option for their use cases.

*Index Terms*—Non-Public Networks, 5G, 3GPP, Industry 4.0.

## I. INTRODUCTION

Connectivity has become a pivotal driver to drive digitalization and product servitization in industrial environments. Industry 4.0 describes the "fourth industrial revolution", which aims at transforming today's factories into intelligently connected production information systems that operate well beyond the physical boundaries of the factory premises. Factories of the future leverage the smart integration of "cyber-physical-systems" and Internet of Things (IoT) solutions in industrial processes [1]. The Fifth Generation (5G) networks can play a key enabling role in this integration, offering programmable technology platforms able to connect a wide variety of devices in an ubiquitous manner [2].

The greatest beneficiaries of a 5G-enabled Industry 4.0 will be the non-telco players, typically operational technology (OT) companies from different vertical sectors such as manufacturing or logistics [3]. OT players, hereinafter referred to as industry verticals, may bring a large number of automation use cases (see Figure 1), most of them with stringent requirements in terms of availability, reliability, low latency, safety, integrity, and positioning with high-accuracy. To meet these requirements in a cost-effective manner, industry verticals may leverage the capabilities provided by 5G technology [4].

| Application Areas \ Use Cases | Motion control | Control-to-control | Mobile control panels with safety | Mobile robots | Remote access and maintenance | Augmented reality | Closed-loop process control | Process monitoring | Plant asset management |
|---|---|---|---|---|---|---|---|---|---|
| Factory automation | X | X | | X | | | | | |
| Process automation | | | | X | | | X | X | X |
| HMIs and Production IT | | | X | | | X | | | |
| Logistics and warehousing | | X | | X | | | | | X |
| Monitoring and maintenance | | | | | X | | | | |

Fig. 1: Application areas and use cases in the industry 4.0 ecosystem. Source: 3GPP TS 22.104 [5].

The 3rd Generation Public Partnership (3GPP) leads the standardization activities in 5G. With the definition of the 5G system architecture in [6], 3GPP provides a reference framework for the deployment and operation of upcoming 5G networks, ensuring global inter-operability and their compliance with IMT-2020 Key Performance Indicators (KPIs). Although the first generation of networks based on the 5G system architecture (3GPP Rel-15) were mainly conceived for public use, the possibility of having 5G networks also deployed for private use has recently raised a lot of interest in the industry community. As a result, their study has recently been included as part of the specifications related to the second phase of 5G networks (3GPP Rel-16 and beyond). This has led to a new classification, whereby 3GPP states that, according to their intended use, networks can be classified into two big categories: Public Land Mobile Networks (PLMNs) and Non-Public Networks (NPNs). On one hand, a PLMN provides network services for public use within a given region, which typically scopes national coverage. A PLMN is operated by a Mobile Network Operator (MNO), who takes the role of PLMN operator. On the other hand, a NPN is intended for the sole use of a private organization, typically an industry vertical. The NPN provides coverage and private network services to devices that are within the vertical's defined premises (e.g. factory, campus). Examples of these devices include sensors, robots, auto-guided vehicles and remote worker's AR-enabled tablets. From here on out, we refer to these devices as NPN devices.

In the industry 4.0 ecosystem, the use of a NPN allows a vertical to have an end-to-end, in-premise 5G network, so that the private traffic can be confined within the boundaries of the defined premises, without the necessity to reach public domain. This is desirable for several reasons, including:

- Quality-Of-Service (QoS) requirements of mission-critical use cases, some of them demanding close-to-zero-ms latency and six nines reliability. The only way to satisfy these challenging requirements is to have dedicated 5G network within the factory, with 5G network functions and service applications as close as possible to the devices and making use of enhanced 3GPP reliability mechanisms, in some cases supported by technologies like Time Sensitive Networking (TSN) and DetNet [7].
- Very high security requirements, met by having strong security credentials and specific authorization mechanisms.
- Isolation from the public domain. This enables protecting the NPN against security attacks or malfunctions (e.g., service outage) in the PLMN.
- Independent network operation for the vertical, allowing him to manage the authentication and authorization of NPN devices, and keep track of their subscription data for accounting and auditing purposes.

However, despite the benefits mentioned above, making NPNs entirely independent of public networks is not always the best solution, either because of business reasons (verticals need to make an initial huge investment, followed by high operational expenditures) or technical reasons (when there is a need to provide NPN devices with connectivity when they are out of NPN coverage). For these cases, integration of the NPN with the PLMN is desirable, so that the MNO can provide device connectivity in out-of-coverage scenarios and reduce entry barriers to verticals. The integration brings open issues that have not been addressed yet in current 3GPP documentation. In view of this, 3GPP SA2 has proposed for Rel-17 a new study item called "Study on enhanced support of Non-Public Networks", which precisely aims to identify these issues and elaborate technical solutions to address them. At the time of writing, this work item has not started yet, although it is planned to begin in the second half of 2019.

In this article, we discuss the use of 5G-enabled NPNs as a means to support industry 4.0 ecosystem. For this end, we will first provide an state-of-the-art of NPN in 3GPP specifications, identifying the work done so far. On the basis of this work, we will identify a number of network implementation options for NPNs that could be relevant for industry 4.0 ecosystem, ranging from NPNs completely separated from a PLMN, to NPNs that are entirely hosted by PLMNs, with some scenarios between these extremes. The selection of one or other option is up to the vertical, who can take this decision based on different criteria that include *i)* service requirements of considered use cases, and *ii)* business-related issues. To help vertical with this decision, we will provide a comparative analysis of the different options, discussing their pros and cons by means of different criteria, including QoS customization, autonomy, isolation, security, service continuity, NPN management and entry barriers for verticals.

The structure of this article is as follows. First, we will provide a overview of the 5G system architecture. Then, we will present the NPN concept in 3GPP ecosystem. Later, we will identify relevant deployment scenarios for NPN, and analyze them based on different criteria. Finally, we will provide some concluding remarks.

## II. OVERVIEW OF THE 5G SYSTEM ARCHITECTURE

The 4G mobile network architecture was designed to meet requirements for conventional mobile broadband services. This architecture, consisting of a large number of coarse-grained network elements connected with point-to-point interfaces, is rather static and too complex to meet the flexibility, elasticity and scalability that are required to efficiently support the wide variety of vertical use cases that may arise in upcoming years. To meet the diversified requirements of these use cases with minimal complexity and costs, 3GPP has defined a completely new system architecture, shown in Fig. 2. In this section, we provide a high-level description of the 5G system architecture. For more details, please see [6]-[8].

The key principles that explain the evolution from the 4G to the new 5G system architecture are the following:

- A converged core network, to support multiple access technologies. The new 5G Core (5GC) supports New Radio (NR), Evolved UTRAN (E-UTRAN) and non-3GPP access (e.g. Wi-Fi, Fixed). NR is the 3GPP air interface technology used in the new 5G radio access network (NG-RAN), consisting of one or more RAN nodes called next-generation NodeB's (gNBs).
- Control User Plane separation (CUPS). Following Software-Defined Networking (SDN) principles, control and user plane functions are separated for completely independent capacity scaling, decoupled technical evolution, and maximum topology flexibility.
- A unified User Plane Function (UPF), with modular forwarding and processing capabilities that can be flexibly programmed by the control plane.
- Compute and storage separation, allowing any network function to store data (e.g. UE and session context) in a centralized database (unstructured data storage function, UDSF), so that data can be shared across multiple instances of this network function. This supports multiple features such as scaling and 1:N resiliency models, making the 5G system more cloud-native.
- Modularization of the architecture design, introducing a set of finer granularity network functions with looser implementation restrictions.
- Service-Based Architecture (SBA), whereby all control plane network functions are connected to a message bus, exposing their functionality to the rest of network functions over service based interfaces. To allow every network function to discover the services offered by other network functions, the network function repository function (NRF) is defined.

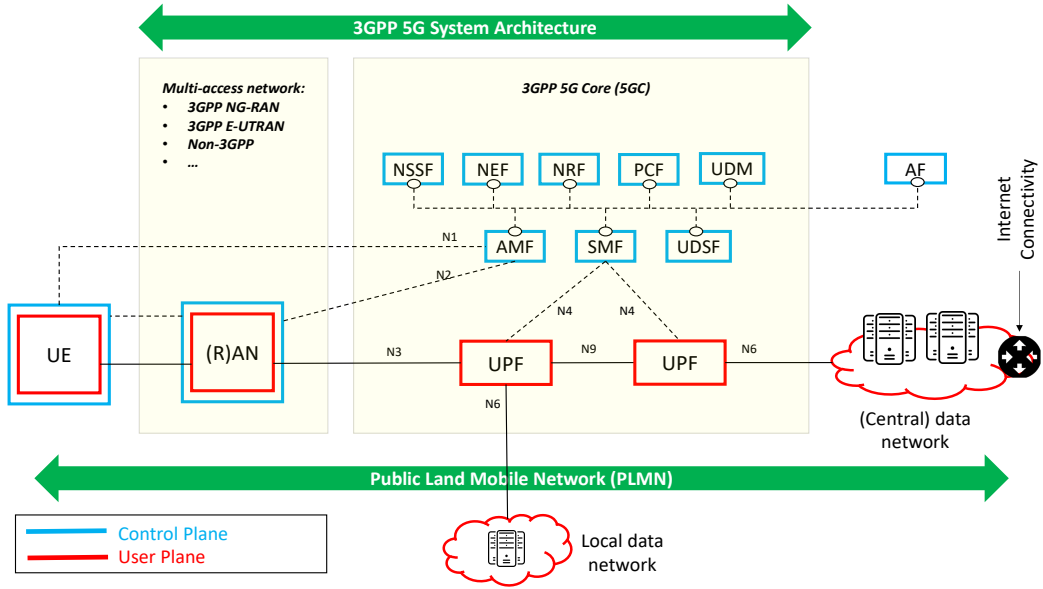

Fig. 2: 3GPP 5G System Architecture

Fig. 2 shows the 3GPP 5G system architecture in the context of the MNO's PLMN. As it can be seen from the figure, 3GPP 5G system only includes Radio Access Network (RAN) and Core Network (CN) domains, but nothing beyond that. This means that data networks connected to the UPF via the N6 reference point are viewed by 3GPP as external network domains. Nonetheless, the role of these data networks is key to ensure effective support of 5G services in an end-to-end manner. In this paper, two types of data networks are considered:

- *Regional data network.* This data network is owned by the MNO, and thus formally belong to the PLMN. It allows the MNO to provide UEs with *i)* internet connectivity, and *ii)* value-added network services, including IP Multimedia Subsystem (IMS) services and non-3GPP L4-L7 services (e.g. firewalling). To host these services, the regional data network consists of one or more high-volume servers with virtualization capabilities.
- *Local area data network.* Unlike the regional data network, a local area data network does not provide internet connectivity, and does not have high compute capacity; indeed, it consists of one or more edge nodes where paradigms like Multi-Access Edge Cloud (MEC) [9] can be applied. These nodes allows hosting delay-sensitive applications (e.g. for closed-loop robot motion control), so that they can be executed as close as possible to the UE. The local data network can be owned by the MNO, or by under the administrative domain of an industry vertical. In the latter case, this data network can belong to a NPN.

In the following sections, we provide a overview of NPNs in 5G scenarios.

## III. 5G-ENABLED NPNS

The standardization work on the use of NPNs in 5G systems is still in its infancy. This is in part due the lack of participation and influence of the OT players into the work progress of the relevant standards development organizations (e.g. 3GPP, ETSI, IETF and ITU). This has resulted in a misalignment between the service requirements in the industrial domain and the technical solutions delivered by the different standardization bodies. A first step to solve this has already been taken in 3GPP, with the definition of two Rel-16 study items: "Communication for Automation in Vertical domains" (3GPP TR 22.804) [10], and "LAN Support in 5G" (3GPP TR 22.821) [11]. In these study items, use cases from different vertical industries have been analyzed, with a special focus on those requiring the use of NPNs. Based on the requirements derived from this analysis, 3GPP has proposed an initial classification for NPNs, whereby NPNs can be divided into two main categories:

- *Stand-alone NPNs*, i.e. NPNs that do not rely on network functions provided by a MNO. A stand-alone NPN is an isolated private network that does not interact with a PLMN; indeed, the NPN and PLMN are deployed on separate network infrastructures.
- *Public network integrated NPN*, i.e. NPNs deployed with the support of a PLMN. Unlike a stand-alone NPN, a public network integrated NPN is hosted (completely or in part) on PLMN infrastructure, relying on some MNO's network functions.

Despite having defined these two NPN categories, 3GPP documents do not provide further elaboration on them. This is the gap we cover in the following subsections, where these categories will be analyzed in detail, identifying some variants that could be found within them. Figures 3 and 4 illustrates these two categories. For the sake of simplicity, we consider that the vertical's defined premises is a factory.

*A. Stand-alone NPN*

A stand-alone NPN is a private network based on the 3GPP 5G system architecture and completely separated from any PLMN. The independence between this NPN and a PLMN is manifested in the following terms: *i)* the use of an unique identifier for the NPN, i.e. NPN ID, entirely independent of the PLMN ID; *ii)* the assignment of private spectrum to the NPN; and *iii)* the full deployment of a 5G system (including RAN and CN) within the logical perimeter of the factory. The fact that the NPN's CN is independent of the PLMN's CN means that subscription data, signalling traffic and user plane flows from NPN devices remain within the boundaries of the factory, and do not crosses PLMN. For this reason, NPN devices are by definition non-public network subscribers.

In order to meet the stringent latency and reliability values required by some use cases, a licensed spectrum is highly preferred for the NPN. This licensed spectrum can be directly obtained from the regulator, or sub-leased from the MNO.

There are some situations where the NPN devices need to access public network services such as voice or internet, while within NPN coverage. In such scenarios, the establishment of a communication path between the NPN and the PLMN is required. As shown in Fig. 3, a firewall can be used for this end. This firewall allows connecting the NPN data network with the PLMN data network. On the one hand, the NPN data network is within the factory, and usually consists of an edge node with MEC capabilities to run vertical-specific service applications. On the other hand, the PLMN data network consists of one or more regional cloud data centers hosting network services provided by the MNO. Note that the NPN and PLMN data networks illustrated in Fig. 3 corresponds to the local area and regional data networks from Fig. 2. Also note that in this scenario, the firewall is a clearly and identifiable demarcation point that allows separation of responsibilities between the NPN operator (i.e. the vertical) and the PLMN operator (i.e. the MNO).

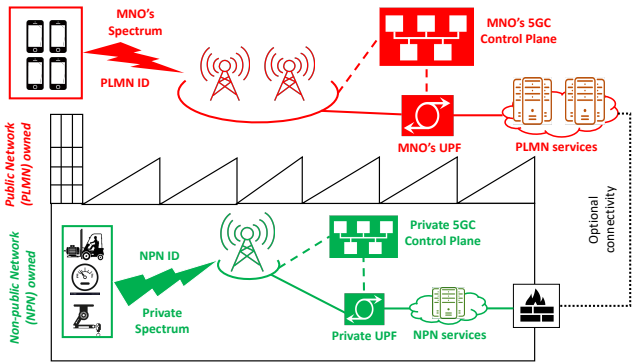

Fig. 3: Stand-alone NPN

*B. Public Network Integrated NPN*

A public network integrated NPN is a private network based on the 3GPP 5G system architecture and deployed in conjunction with a PLMN. This category assumes the NPN consists of one public sub-network and one or more private sub-networks. On one hand, the public sub-network contains PLMN provided network functions. These functions are under the MNO's administrative domain, and usually deployed out of the factory. On the other hand, a private sub-network includes network functions that remain segregated from the PLMN, and that are allocated inside the factory. The deployment of public and private sub-networks in a public network integrated NPN can vary depending on the considered use case. In this paper, four deployment scenarios have been identified in this respect:

- Shared RAN, with MORAN[12]-based approach (scenario B.1, Fig. 4a): the NPN and PLMN have different IDs, segregated spectrum bands, and independent CNs. As seen, this scenario is quite similar to an stand-alone NPN, with NPN devices being non-public network subscribers. The novelty that this scenario brings is that the RAN segment of the NPN is partially shared with the PLMN. This means that some functions of the RAN nodes serving NPN devices within the factory can be provided by the PLMN. These functions are shared between the NPN and the PLMN, and thus define the public sub-network of the NPN. The rest of RAN functions remain segregated, and thus taken part in a private sub-network of the NPN.
- Shared RAN, with MOCN[12]-based approach (scenario B.2, Fig. 4b): this scenario is similar to B.1, with the exception that the NPN and PLMN also share the spectrum. As it can be seen from the figure, this spectrum is public and owned by the MNO.
- Shared RAN and shared CN control plane (scenario B.3, Fig. 4c): in this scenario, the only part of the NPN that remains entirely separate from the PLMN is the CN user plane. The CN control plane is provided by the PLMN, which means the *i)* network control tasks in the NPN are performed in the MNO's administrative domain, and *ii)* NPN devices are by definition public network subscribers. In this scenario, segregation of non-public and public traffic portions can be achieved by means of 3GPP-defined mechanisms, including network slicing.
- Shared RAN and CN (scenario B.4, see Fig. 4d): the NPN is entirely hosted by the PLMN. This means that both public and non-public traffic portions are external to the factory, with all data flows routed towards the PLMN via the shared RAN node. However, to guarantee the separation and independence of both portions, these need to be treated as part of completely different networks. To enforce the needed segregation, slicing can also be used.

As in the case with the stand-alone NPN, a firewall installed in the outer edge of the factory allow connectivity between public and private domains. The presence of this firewall is optional for the scenarios B.1, B.2 and B.3, and it is only

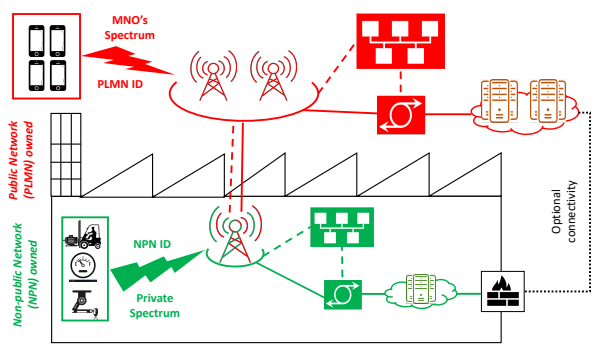
(a) Scenario B.1: The NPN has a dedicated (non-MNO-owned) spectrum, but shares (part) of the RAN functionality with the PLMN.

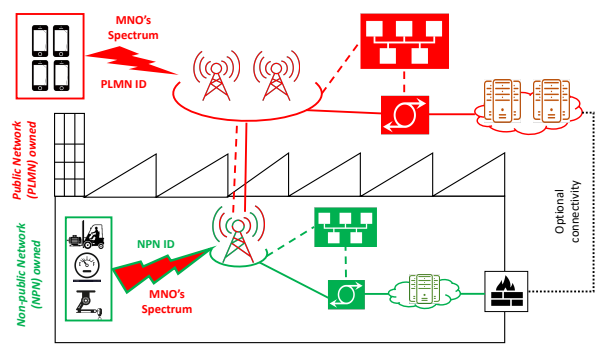
(b) Scenario B.2: The NPN shares spectrum bands and (part of) the RAN node functionality with the PLMN.

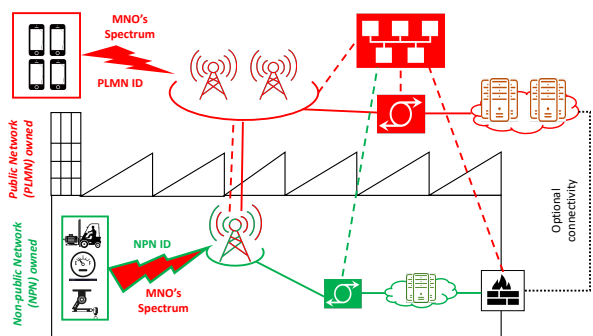
(c) Scenario B.3: The NPN has a dedicated CN user plane. The rest of the 5G system (CN control plane and RAN) is shared with the PLMN.

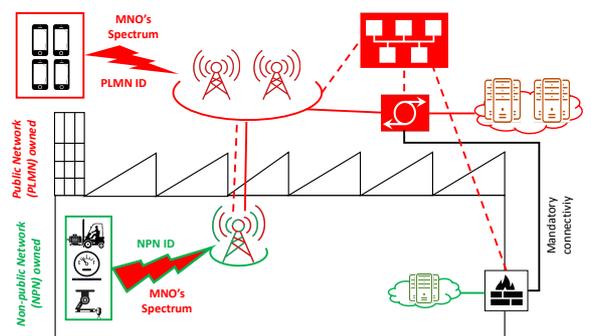
(d) Scenario B.4: The NPN shares all the 5G system components with the PLMN.

Fig. 4: Scenarios for the public network integrated NPN category

required when NPN devices want to consume public network services. It is however mandatory for the scenario B.4, since the firewall is the only way to allow these devices to access NPN services through the PLMN. For this end, the firewall connects PLMN provided UPF with the NPN data network.

## IV. ANALYSIS ON NPN ATTRIBUTES

This section focuses on the attributes that are relevant for the 3GPP-defined NPNs, analyzing their implications for the different scenarios discussed in Section III. The degree of compliance with these attributes should be considered by an industry vertical when assessing the suitability of an NPN deployment scenario for any planned industry 4.0 use case.

### A. QoS customization

It describes the ability to flexibly configure the parameters governing the behavior of a NPN, in such a way that the NPN can satisfy the specific requirements of targeted use cases. These requirements include coverage profiles (e.g. indoor), traffic patterns (e.g. uplink/downlink frame structures) and KPI values (e.g. throughput, latency and jitter, reliability values) that are dependent of the use case under consideration, and quite different from those typically considered in PLMNs. The more independent of a PLMN a NPN is, the more flexibility in NPN parameters setting is allowed, which naturally leads to a use case-tailored NPN configuration.

### B. Autonomy

It is the ability to guarantee the normal operation of the NPN, regardless of any unexpected event (e.g. security failure, performance degradation) occurred in the PLMN.

### C. Isolation

Isolation in NPN scenarios is the ability to make non-public traffic portion independent of any other traffic portion flowing in the PLMN infrastructure. This independence shall be assessed *i)* in an end-to-end manner, from the device to the data network; and *ii)* across the different networking planes, including user, control and management planes. In many deployment scenarios such as those considered in the public integrated NPN category, the NPN and the PLMN share (part of) the same infrastructure resources. It is therefore necessary to consider possible forms of isolation for those scenarios, according to their specificities. Despite their differences, all these forms of isolation need to converge into the following two principles:

- 3GPP network functions from the NPN and the PLMN shall be deployed separate from each other. This separation can be enforced not only at the physical level, but also at the logical level. The latter is particularly relevant for scenarios B.3 and B.4, with high levels of sharing between the two networks. In these scenarios, isolation can be guaranteed through the application of NFV paradigm (i.e. deploying 3GPP functions as virtual network functions) and the corresponding protection mechanisms. This protection shall be mostly focused on how resource sharing is applied, avoiding situations of resource starvation when a certain function sharing a virtualized infrastructure of any nature gets overloaded, depriving other functions of needed resources.

| Attribute | Stand-alone NPN | Public network integrated NPN | | | |
|---|---|---|---|---|---|
| | | Scenario B.1 | Scenario B.2 | Scenario B.3 | Scenario B.4 |
| *QoS customization* | Full customization | Full customization | High customization | Partial customization | No customization |
| *Autonomy* | Full autonomy | High autonomy | PLMN failure in the RAN node most likely leads to NPN failure. | PLMN failure most likely leads to NPN failure | PLMN failure leads to NPN failure |
| *Isolation* | Nothing is shared. NPN device subscription data and user plane flows confined within the factory. | RAN node sharing. NPN device subscription data and user plane flows confined within the factory. | RAN node sharing. NPN device subscription data and user plane flows confined within the factory. | Only UPF is dedicated. NPN device subscription data stored in the PLMN. User plane flows confined within the factory. | All NPN's network functions hosted by the PLMN. Subscription data and user plane flows from NPN devices leave the factory. |
| *Security* | High security, due to full isolation between NPN and PLMN. | High security, due to full isolation between PLMN and NPN. | Dependent on *i)* PLMN-defined security mechanisms enforced at the RAN node, and *ii)* vertical-specific intra-NPN security mechanisms. | Dependent on *i)* PLMN-defined security mechanisms enforced at the RAN node & CN control plane, and *ii)* vertical-specific security mechanisms enforced at the UPF. | Completely dependent on PLMN-defined mechanisms. |
| *Service continuity* | Dependent on the MNO. No solutions agreed so far in 3GPP. | Dependent on the MNO. No solutions agreed so far in 3GPP. | Requires the use of the N3IWF. This can be complemented with UE dual radio support. | Should be easy, according to 3GPP defined roaming mechanisms. | Always guaranteed, as long as roam agreement is signed between the MNO and the vertical. |
| *NPN management for verticals* | Full control | Full control | High level of control, although some deployment changes might require MNO support | Limited control. Configuration settings on the UPF could be used for modifications on SLA requirements. | Very limited control. mostly focused on performance assurance and fault supervision activities. |
| *Entry barriers* | Very high | Very high | Medium | Low | Very low |

TABLE I: Analysis of NPN features for different deployment scenarios

- Data of public and non-public network subscribers need to be segregated and processed separately, in order to safeguard necessary privacy of the vertical and the MNO. To achieve this, it should be sought, to the extent possible, avoid transmitting and storing private data outside the boundaries of the vertical's defined premises.

### D. Security

Guaranteeing security in industrial scenarios requires that NPN communications provide full confidentiality and integrity, in particular when traversing PLMN paths shared with other traffic flows, which will be most, if not all, in practically any feasible scenario. This requires:

- The use of well-known network security techniques, to ensure the required confidentiality and integrity, and poses the challenge of how cryptographic material is distributed in an acceptable way to the different deployed 3GPP network functions. Such an acceptable way implies it is trustworthy, so no element impersonation can happen, and verifiable, so identities can be securely verified by the communicating parties. For more details, see [13].
- Segregating the control plane and the management plane functions for the NPN and PLMN, to ensure that the vertical is only able to access network functions specific to the NPN (e.g. for configuration, accounting and/or auditing purposes), and unable to access other similar network functions specific to the MNO's PLMN.

### E. Service continuity

It is the ability o to provide zero-time service interruption when the NPN devices moves between the NPN and the PLMN, and viceversa. Service continuity assumes that PLMN is able to provide seamless connectivity to a device when leaving NPN coverage, either due to a temporal outage in the NPN, or simply because the device moves between two NPNs placed in different locations, although serving the same vertical, e.g. two factories administrated by the same vertical. To avoid service interruption in this type of situations, interworking mechanisms scoping signaling (e.g. automatic network selection) and security (e.g. certificates for device authentication and identification, and for access authorization) should be designed. Apart from well-studied roaming procedures, novel mechanisms based on the use of Non-3GPP Interworking Function (N3IWF) [6] are being explored in 3GPP specifications for this end. The N3WIF, deployed at the NPN (and the PLMN), performs a gateway-like functionality that allows handing over sessions from the NPN to the PLMN

(and viceversa) when UE moves between both networks. This N3WIF-like gateway solution can be complemented with UE dual radio support mechanisms, as described in [14].

*F. NPN management for verticals*

It refers to how much control the vertical can take to freely manage the NPN and its network functions. The more control the vertical company has, the better it can adapt the behavior of the NPN to the specific needs of the served use cases, in terms of performance, functionality and scalability. This control can be exercised through *i)* the administration of specific policies; *ii)* the execution of performance assurance and fault supervision activities; and *iii)* the life cycle management of network functions and service applications, particularly relevant in NFV environments.

*G. Entry barriers for verticals*

A business KPI relevant for any industry vertical is the cost of having an NPN up and running. To estimate the amount of money a vertical shall invest for this purpose, a wide variety of cost sources should be assessed, including *i)* spectrum acquisition; *ii)* purchase/rental and maintenance of the compute, storage and networking hardware within the factory; *iii)* purchase/rental and maintenance of software images executing CN functionality, if virtualized; and *iv)* operational expenditure for a management and orchestration solution.

Table I provides an comparative analysis of the different attributes and the implications these have for the different NPN deployment scenarios. As it can be seen, deployments close to stand-alone NPN and scenario B.1 make the NPN entirely independent in terms of performance, management and security; however, they may require significant investment from vertical industry side and might introduce some interworking issues with the PLMN, which hinders service continuity in mobility scenarios. These type of NPN deployment scenarios are ideal to support mission-critical, delay-sensitive industrial industrial use cases demanding full isolation guarantees, and where participating devices are rather static. On the other hand, deployments close to scenarios B.3 and B.4 makes NPN more dependent on PLMN behaviour. This facilitate the interaction between the two networks and lower entry barriers for verticals, at the costs of making NPNs less isolated in terms of performance and management. The selection of one or another deployment scenario depends on the cost-benefit ratio from the vertical's viewpoint, considering the requirements of the targeted use case.

## V. CONCLUSIONS AND FUTURE OUTLOOK

In this paper we have described a number of deployment options for NPNs in the industry 4.0, based on 3GPP 5G specifications. These range from NPNs completely separated from a PLMN (stand-alone NPNs), to NPNs that are entirely hosted by the PLMN (scenario B.4). We also have provided a comparative analysis of the different options, based on different criteria. The outcome of this analysis may be useful for those verticals interested in NPNs, helping them to decide what is the best deployment option for them, according to their specific service needs and considering the effort they are willing to invest in designing, deploying and operating NPNs.

This paper provides guidelines that can be used as starting point for further progress in NPN standardization in 3GPP 5G systems. Much of the work that needs to be undertaken in the future includes the study on the applicability of network slicing in scenarios B.3 and B.4, and the study on interfaces to enable interworking and seamless handovers between NPNs and PLMNs. Apart from these technical issues, other aspects should also be considered and explored, including:

- Regulatory aspects, most of them related to the spectrum.
- New business models. Indeed, the integration of NPN and PLMN enables a synergistic relation whereby industry verticals have incentives to invest in on-premise 5G evolved infrastructure, which can then be offered "as a service" to one or more MNOs to allow them to expand their service footprint at a reduced cost.

ACKNOWLEDGMENT

This work is supported by the H2020 European Projects 5G-VINNI (grant agreement No. 815279) and BOOST 4.0 (grant agreement No. 780732).